\documentclass[onecolumn]{aa}
\usepackage{graphicx}
\begin{document}
   \title{The Coma Cluster at $\gamma$-ray energies: multifrequency constraints}

   \author{A. Reimer
          \inst{1}
          \and
          O. Reimer\inst{1}
          \and
          R. Schlickeiser\inst{1}
          \and
          A. Iyudin\inst{2}
          }

   \offprints{afm@tp4.rub.de}

   \institute{Institut f\"ur Theoretische Physik IV,
              Ruhr-Universit\"at Bochum,
              D-44780 Bochum, Germany
        \and
	      Max-Planck Institut f\"ur extraterrestrische Physik,
              Gie{\ss}enbachstra{\ss}e,
              D-85740 Garching, Germany
	     }

   \date{Received March 19, 2004; accepted June 1, 2004}

   \abstract{The Coma cluster exhibits evidence for a high-energetic non-thermal
particle population. At frequencies $>1$GHz recent radio halo observations confirm a
significant spectral steepening of the volume-integrated emission. We calculate the
volume-averaged high-energy spectrum due to inverse Compton scattering off the CMB
radiation field and non-thermal bremsstrahlung according to an exponential cutoff
in the electron spectrum as deduced from the radio observations. The synchrotron radiation
from secondary pairs, created from the decay of charged mesons produced in hadronic $pp$-interactions,
is found to set significant constraints on the energy content of relativistic hadrons in Coma.
This limits the maximum flux at high energies. Our findings support a low ratio of relativistic
hadron to thermal energy density. Predictions for Coma's high energy emission are discussed in
the light of current and expected abilities of upcoming $\gamma$-ray instruments.

    }

   \titlerunning{The Coma Cluster at $\gamma$-ray energies}
   \maketitle
%

\section{Introduction}
Clusters of galaxies are conglomerates of a large number of galaxies
that are gravitationally bound and confining a large fraction of the mass in the universe.
One of the controversially
discussed properties of clusters of galaxies are their non-thermal components which include
cosmic rays as well as turbulence and non-regular magnetic fields.
The non-thermal pressure has an important impact on the evolution of galaxy clusters.
To achieve a better understanding of these components is mainly driven by the importance
of the non-thermal pressure in the evolution of galaxy clusters.
In this paper we discuss only cosmic rays as the easiest testable one of all non-thermal components.
Several mechanisms have been proposed that lead to
relativistic particles in the intracluster medium (ICM), as e.g. particle
acceleration during their formation and
evolution, in e.g. merger shocks
(\cite{Takizawa2000},
\cite{GabiciNonPL}, \cite{Berrington2003}, \cite{Miniati2001a}),
accretion shocks (\cite{Cola1998}),
intergalactic termination shocks from the
winds of the galaxies (\cite{Voelk1996}), or
reacceleration of injected mildly relativistic particles
from powerful cluster members (\cite{Ensslin1997}).
Indeed, the detection of synchrotron radiation from cluster radio halos and relics signals
the existence of relativistic electrons (e$^-$) in the intracluster medium (ICM).
The most prominent cluster which possesses a radio halo is the Coma cluster
(Abell 1656), located at redshift z=0.0232 (\cite{Struble1991})
(corresponding to $\sim90$~Mpc for H$_0$=75~km~s$^{-1}$~Mpc$^{-1}$).
At frequencies below $\sim 1$~GHz observed volume-integrated fluxes are satisfactorily fitted by
a pure power law.
Observations at larger frequencies gave evidence that a significant spectral steepening of the
integrated emission occurs in Coma's radio halo (\cite{Schlickeiser1987}), recently confirmed by
\cite{Thierbach2003} (TH03). A hard X-ray (HXR) excess has been detected by the Rossi X-ray Timing Explorer
(RXTE) (\cite{Rephaeli1999}) and BeppoSAX (\cite{Fusco1999}). Coma is one of the few clusters
where also an EUV excess is conclusively established (\cite{Bowyer1998}, \cite{Lieu1999}).
The spread of the soft X-ray to EUV emission is, however, still debated
(\cite{Kaastra2003}), and may possibly extend up to the turnaround radius.
We restrict here our considerations to the size of the radio halo
($\sim$10'$\times$30'; TH03).
By comparing the radio synchrotron spectrum with this excess radiation, interpreted
as Inverse Compton (IC) scattering off photons from the cosmic microwave background (CMB) by the same
e$^-$ population, volume-averaged magnetic
fields of $B = 0.1\ldots0.3\mu$G have been deduced (\cite{Fusco1999}, \cite{Rephaeli1999}).
Faraday rotation measurements gave
$B\sim 2\ldots10\mu$G (\cite{Kim1990}, \cite{Clarke2001}).

Many models for non-thermal radiation from the Coma cluster predict significant emission
at $\geq 100$~MeV due to both, relativistic e$^-$ and ions (\cite{Sarazin1999a}, \cite{Atoyan2000},
\cite{Miniati2003}, \cite{Gabici2004}). They often assume a straight power
law to $\sim 10^{6\ldots7}$MeV for the e$^-$ population responsible for the
dominating synchrotron component.
Here we investigate the consequences of the decline in the e$^-$ spectrum at $\sim 10^4$MeV as
deduced from recent radio observations of Coma C for the expected high energy flux.
In Sect.~2 the IC and non-thermal bremsstrahlung spectrum from the steepening e$^-$
distribution is calculated. In Sect.~3 we derive limits imposed by the broadband observations
for the $\pi^0$-decay $\gamma$-ray component
including its secondary pair initiated radiation.
Finally we discuss Coma's detectability with current/future $\gamma$-ray instruments
such as INTEGRAL, GLAST-LAT and the new Imaging Air Cherenkov Telescopes (IACTS).


\section{Non-thermal electron spectrum and radiation}

\begin{figure}
\resizebox{\hsize}{!}{\includegraphics{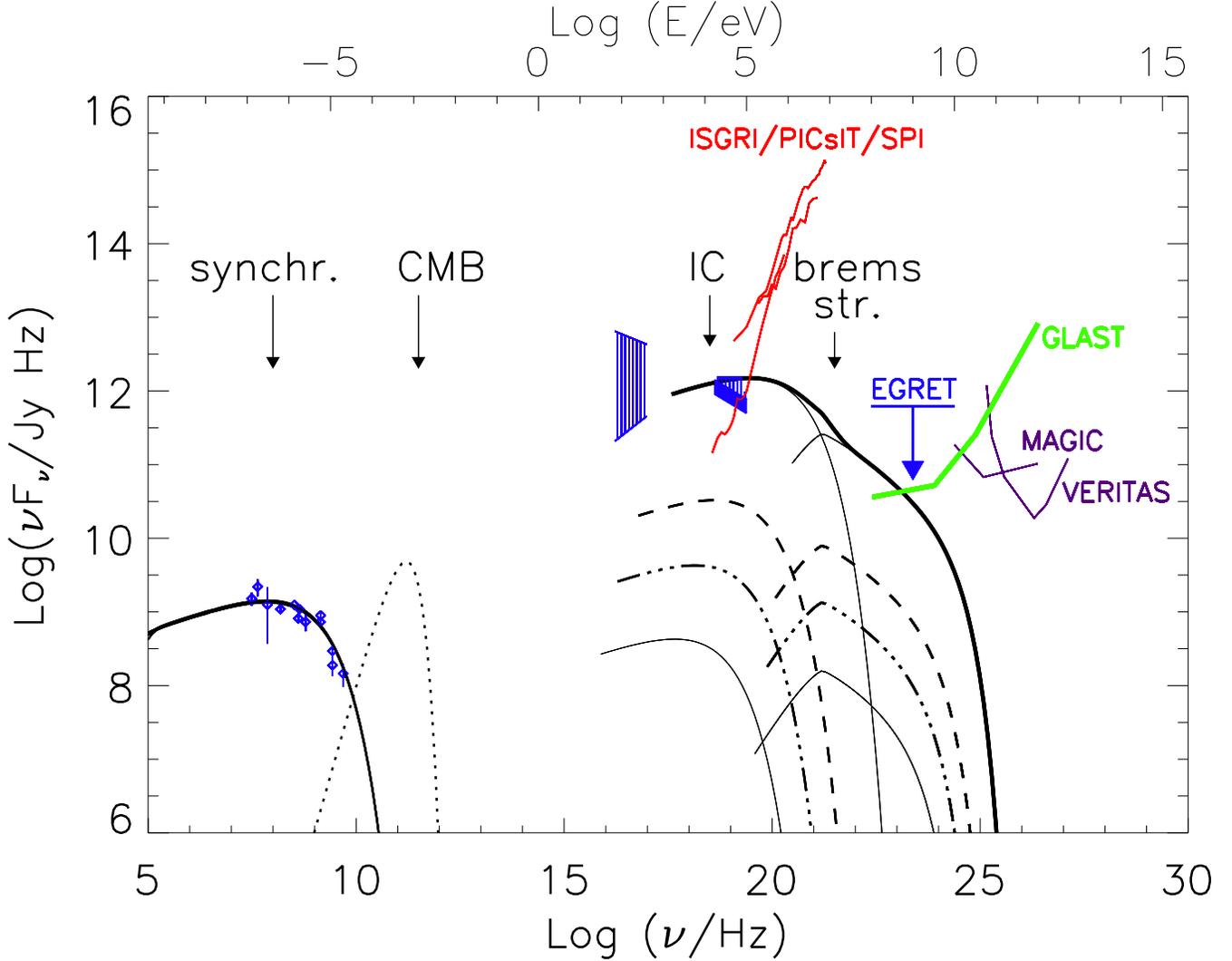}}
\caption{Broad band continuum spectrum of Coma. The radio data and the best-fit
spectrum at source (corrected for self-absorption) are taken from TH03
(not corrected for the thermal Sunyaev-Zeldovich effect). The dotted line
represents the CMB field corrected
for the thermal Sunyaev-Zeldovich effect using a y-parameter of $0.75\cdot10^{-4}$ (\cite{Ensslin2002}).
The IC and non-thermal bremsstrahlung fluxes are calculated for
field strengths $B=0.1$, 0.68, 1.9, 6$\mu$G (from upper to lower
curves) using an exponential e$^-$ distribution (with adjusted $Q_0$, $p_c$ to fit the radio data),
and $n_i=10^{-3}$cm$^{-3}$.
They are extended to lower energies assuming the synchrotron spectrum follows a power law down to at least $10^{-9}$eV.
The hatched regions in the X-ray domain represent the data from PDS/BeppoSAX (\cite{Fusco1999}),
HEXTE/RXTE (\cite{Rephaeli1999}) and EUVE (\cite{Lieu1999}). The RXTE and EUVE fluxes are integrated within a ring of 21' and 18',
respectively, while the PDS data include only fluxes within 8' from the cluster center.}
\label{fig1}
\end{figure}

The volume-integrated
radio emission from the radio halo has been studied in detail by e.g. \cite{Schlickeiser1987},
\cite{Kim1990}, \cite{Venturi1990}, \cite{Giovannini1993}, \cite{Deiss1997}, TH03.
Fig.~\ref{fig1} shows the volume-integrated radio continuum spectrum of the diffuse radio halo
source Coma C as published in TH03 with the best fit model.
TH03 confirmed the findings of \cite{Schlickeiser1987} that
among the three basic models for cluster halos
(primary electron model: \cite{Jaffe1977}, \cite{Rephaeli1979};
secondary electron model: e.g. \cite{Dennison1980}; in-situ acceleration model: \cite{Jaffe1977}, \cite{Roland1981},
\cite{Schlickeiser1987})
the in-situ acceleration model fits the observed exponential
steepening of the synchrotron spectrum best. This model, though discussed critically by \cite{Petrosian2001},
considers shock wave and resonant diffusion acceleration
out of a thermal pool of particles where radiation losses and particle escape has been taken into account.
A secondary origin of the radio halo has been pushed forward by many authors (e.g. \cite{Dennison1980},
\cite{BlasiCola99}, \cite{Dolag2000}, \cite{Atoyan2000}, \cite{Blasi2001},
\cite{Miniati2001b}). Recently, however, arguments are given which suggest that secondary pairs
as the underlying particle population of the radio halo emission are problematic
(\cite{Brunetti2003}, \cite{Kuo2004}).
Along these lines \cite{Brunetti2004} found that the observations of non-thermal radiation of galaxy clusters
are only reproducable within the picture of particle acceleration through cluster merger generated
Alv\'en waves,
if the fraction of relativistic hadrons in the ICM is small (5-10\%). This hadron content is insufficient
to reproduce the radio halo from secondary pairs (see below).
Curved spectra are also possible at an energy where losses balance the acceleration rate
if the acceleration time decreases more slowly than the loss time.
In the following we therefore consider an exponential shape of the e$^-$ spectral
distribution, suitable to explain the volume-averaged synchrotron spectrum
,
irrespective of its formation mechanism. This rather phenomenological ansatz will not shed light onto the
mechanisms responsible for the formation of the e$^-$ distribution, however,
leads to model-independent constraints for the high-energy component arising from
this leptonic particle population.

We fit the radio flux density with a power law synchrotron spectrum
extended by an exponential cutoff:
$$
I_{\rm syn} (\nu) \propto \nu^{(3-s)/2} \exp{(-\sqrt{\nu/\nu_s})}
$$
with $s = 4.6$ and $\nu_s = 0.44$~GHz (TH03). Synchrotron-self absorption will affect
the synchrotron spectrum at low energies. For an estimated path length of $\sim$280~kpc
through the cluster (TH03)
we found the turnover frequencies at $\sim$ a few 100~kHz
(see Fig.~\ref{fig1}).
Below $\sim$ 0.2~MHz free-free absorption in the disk of the Milky Way
suppresses the radio intensity from Coma observed at Earth.
In addition,
the Razin-Tsytovich effect causes the radio spectrum
to decline rapidly to low frequencies from $\nu_R \la 20 (n_e/B_G)$~Hz ($n_e$ is the e$^-$ density in cm$^{-3}$, and
$B_G$ is the magnetic field in Gauss) with turning point between $0.01\ldots0.6$~MHz for the expected field
strengths in Coma.
Turnovers below MHz-frequencies are, however, not possible to detect with ground based radio observatories
due to ionospheric effects, and have to await future space-based low-frequency observatories, for example a
radio observatory on the Moon.

For a given magnetic field $B$ the corresponding volume-integrated e$^-$ spectrum
$$
\eta(p) = Q_0 p^{-s} \exp{(-p/p_c)}
$$
(\cite{Schlickeiser1987}) with normalization $Q_0$ and cutoff e$^-$ momentum $p_c$ can then be determined.
Values for $p_c$ considering magnetic field strengths of 0.1\ldots6$\mu$G lie at $2153\ldots16667$MeV/c.
IC scattering off CMB photons with photon energies
$\bar\epsilon_{\rm CMB}\simeq 6\cdot 10^{-4}$ eV ($T_{\rm CMB}=2.7$~K) by these e$^-$ is
therefore restricted to the Thomson regime.
In the $\delta$-approximation the intensity of the IC scattered radiation as a function of
photon energy $E_\gamma$ can be analytically expressed if the target photon density distribution $n(\epsilon)$
is sufficiently
peaked so that $E_c = (p_c/m_e c)^2\epsilon \approx (p_c/m_e c)^2\bar\epsilon$:
\begin{eqnarray*}
I_C(E_\gamma) & = & \frac{c \sigma_T E_\gamma}{4\pi d_L^2} \int_0^{\infty} d\epsilon n(\epsilon)
\int_{p_{\rm min}}^{\infty} dp p^2 \eta(p) \delta\left(E_\gamma-(\frac{p}{m_e c})^2\epsilon\right)=\\ \nonumber
    & = & \frac{c \sigma_T Q_0}{8\pi d_L^2} \left(kT_{\rm CMB}\right)^{(3+s)/2} \Gamma\left(\frac{s+3}{2}\right) \xi\left(\frac{s+3}{2}\right) \nonumber \\
    & & \left(m_e c\right)^{3-s} E_\gamma^{(3-s)/2} \exp{\left(-\sqrt{E_\gamma/E_c}\right)} \nonumber
\end{eqnarray*}
with $\Gamma$ the Gamma function, $\xi$ is Riemann's zeta function, $m_e c^2$ the e$^-$ rest mass,
$\sigma_T=6.65\cdot 10^{-25}$cm$^2$,
$d_L$ Coma's luminosity distance and
$n(\epsilon)$ is the CMB photon density.
Fig.~\ref{fig1} shows the resulting IC spectra for an equipartition magnetic field ranging between $0.68\ldots1.9\mu$G
(TH03), for
the central magnetic field in Coma C ($B\sim 6\mu$G: \cite{Feretti1995})
and for $B=0.1\mu$G appropriate to explain the HXR excess emission.

The non-thermal volume-averaged bremsstrahlung intensity using the primordial $^4$He mass fraction
of 0.24
$$
I_B(E_\gamma) = \frac{1.18 n_i c E_\gamma}{4\pi d_L^2} \int_{\max(E_\gamma m_e c, p_{\rm min})}^{p_{\rm max}} dp p^2 \eta(p) \frac{d\sigma}{dE_\gamma}
$$
is calculated in the relativistic limit
using the differential cross section
from \cite{Blumenthal1970}. For a mean gas density $n_i\sim10^{-3}$cm$^{-3}$ in Coma
we find the compound IC and bremsstrahlung spectrum from the observed synchrotron-emitting
e$^-$ distribution always below the EGRET upper limit (\cite{Reimer2003}).
The steepening of the e$^-$ spectrum at $10^{3\ldots4}$MeV causes
the IC component to decline at $\sim$1-10~MeV, non-thermal bremsstrahlung
dominates till its decline at a few GeV. This is in contrast to works where the
primary e$^-$ spectrum extends to several $10^7$MeV (e.g. \cite{Atoyan2000}, \cite{Miniati2003}).
Here the volume-averaged IC and bremsstrahlung spectra extend to GeV-TeV energies and may
dominate the $\gamma$-ray domain, depending on the strength of the $\pi^0$-decay component.

\section{Hadronic cosmic ray -- gas interactions}

Interactions between cosmic ray protons and nucleons of the ICM gas component are very rare
and occur on average once in a Hubble time in Coma.
The number of collisions is usually time-dependent,
in particular higher than average soon after a (e.g. merger) shock has started to develop,
due to newly injected particles.
In case of a high cosmic ray hadron content $\gamma$-rays from the decay of $\pi^0$
are expected to determine the energy range $>1$~GeV.
Additionally, radiation from the secondary pairs, generated through the decay of
charged mesons that are produced by hadronic $pp$-collisions, is expected to contribute
to the overall broad band spectrum.
The short cooling time scales of those pairs radiating in the GHz and hard X-ray to $\gamma$-ray band
leads to quasi-stationary pair populations at these energies on a very short time scale.
As a consequence a direct relation between the $\pi^0$-decay $\gamma$-ray spectrum and the
radiation spectrum from the (high energy) secondary pairs is expected.

\begin{figure}
\resizebox{\hsize}{!}{\includegraphics{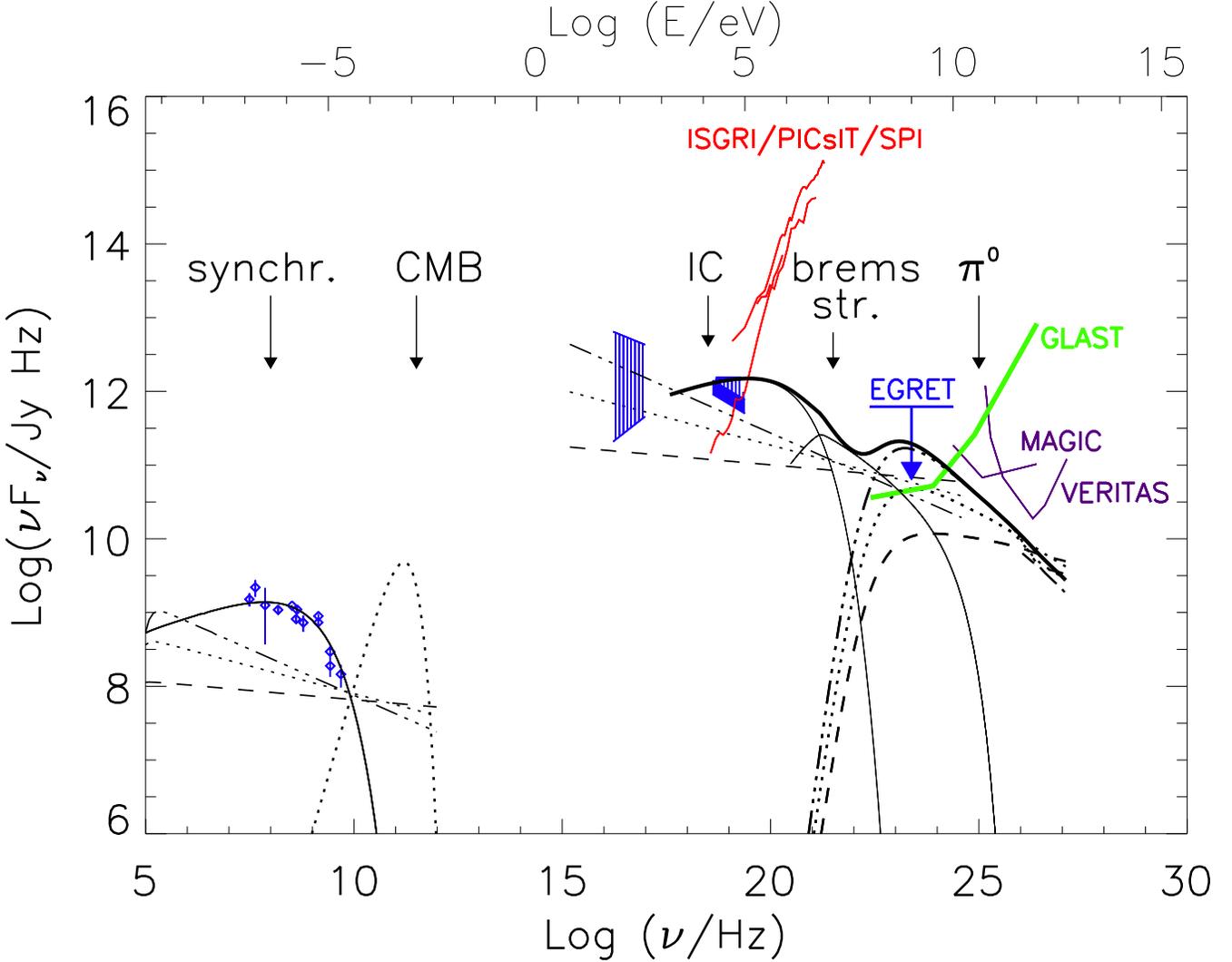}}
\caption{Same as Fig.~1 but
the IC and non-thermal bremsstrahlung fluxes are shown only for a
field strength $B=0.1\mu$G.
The $\pi^0$-decay $\gamma$-ray spectra (most right curves) are calculated for a $\alpha_p=2.1$ (dashed line),
2.3 (dotted line), 2.5 (dashed-dotted line) proton spectrum and
the normalization of the particle spectra are adjusted to avoid violating the EGRET upper limit
as well as the integral fluxes in the HXR and radio domain
(see text). The required relativistic proton energy densities
are 3\%, 8\% and 28\%
of the thermal energy content
for $\alpha_p=2.1$, 2.3 and 2.5, respectively.
The corresponding IC and synchrotron fluxes
are shown as dashed/dotted lines.}
\label{fig2}
\end{figure}

The spectral index of Coma's putative relativistic proton
distribution nor its normalization has yet been determined
observationally. Nevertheless some plausible arguments can be found to limit the parameter space.
Because cosmic ray protons are stored efficiently in galaxy clusters for cosmological times
(\cite{Voelk1996}, \cite{Berezinsky1997}),
the radiation from the secondary pairs reflects the injected proton spectrum, and
the global proton spectrum should be not significantly
different from the injected
one if uniform injection throughout the cluster is assumed.
The structure formation shock scenario gives injection spectral indices of
$\alpha_p=2.0\ldots2.5$ for strong shocks (\cite{Miniati2003}).
For merger shocks plunging into the cluster body from the periphery $\alpha_p$ can evolve from $2\ldots5$
(\cite{Berrington2003}).
The normalization of the proton component is limited by three constraints: Firstly,
the $\pi^0$-decay $\gamma$-rays
must not be overproduced to violate the EGRET upper limit.
Secondly, IC scattering off CMB photons by the secondary pair ($e^\pm$) population produced
in $pp$-interactions leads to a further radiation component
that covers the energy range from $\sim 5$~eV (corresponding to
e$^\pm$ of energy $\sim50$~MeV) to a few GeV.
This component is constrained by the HXR flux and EGRET upper limit.
The expected non-thermal bremsstrahlung from these secondary e$^\pm$ lies always below the corresponding
IC flux level.
And lastly, these secondary e$^\pm$ also emit synchrotron photons,
and this leads to a constraint imposed by the radio observations.
Fig.~\ref{fig2} shows the resulting stationary $\gamma$-ray spectra for
$\alpha_p=2.1, 2.3$ and $2.5$, calculated using the formalism given
in \cite{Pfrommer2003} for the $\pi^0$-decay $\gamma$-ray production and secondary pair production.
We limited the proton spectrum,
assumed to be uniformly injected throughout the cluster,
to $10^6$~GeV since higher energetic protons are difficult to confine within the cluster size
(\cite{Cola1998}).
The use of gas and proton density profiles as applicated in e.g. \cite{Blasi99}
instead of the volume-averaged parameters leads to only minor changes
in the $\pi^0$-decay $\gamma$-ray intensity for the here considered volume of Coma
(with an effective radius of $\sim 330$~kpc).
Above $\sim 1$~TeV photon absorption due to
photon-photon pair production in the cosmic infrared-to-optical background radiation field must be
taken into account. For this correction we used the background models in
\cite{Aharonian2001}.

Proton energy densities $u_p$ are calculated in the following from the
proton spectrum above the threshold for hadronic $pp$-collisions, and are compared to Coma's
thermal energy density $u_{\rm therm}\approx 3.8\cdot 10^{-11}$erg/cm$^{-3}$
(for $kT_e=8.2$~keV, a thermal e$^-$ density of $10^{-3}$cm$^{-3}$ and a $^4$He mass fraction
of 0.24).
The synchrotron flux in the MHz-to-GHz regime from the secondary pairs is dependent on
$u_p$ and $\alpha_p$ as well as on $B$.

For $\alpha_p\sim 2.4$ and $X_p\equiv u_p/u_{\rm therm}\sim 20\%$ the radio data
are explainable by synchrotron emission from secondary e$^\pm$ in a volume-averaged
magnetic field of $0.15\mu$G
if the steepening of the radio spectrum
at high frequencies is disregarded,
in agreement with \cite{BlasiCola99}, \cite{Dolag2000}.
If the steepening of the $>1$~GHz radio data is taken into account, obviously the synchrotron flux from the secondary pairs
must lie below the GHz-radio observations. In fact, we find that these high frequency radio data
place the most stringent constraint on the proton energy content in the Coma Cluster.
The resulting upper limits for the relativistic hadronic energy density of
$X_p<3\%\ldots0.009\%$,
$X_p<8\%\ldots0.01\%$
and  $X_p<28\%\ldots0.07\%$ (assuming $B=0.1\ldots2\mu$G) for
$\alpha_p=2.1$, 2.3 and 2.5, respectively, are significantly lower as used
in structure formation triggered acceleration scenarios. For example, the model of \cite{Miniati2003} required $\sim 34\%$
of the thermal energy in form of cosmic ray ions for $B=0.15\mu$G, and $\sim 4\%$ for $B=0.5\mu$G with
a proton spectrum $\alpha_p\sim 2$ to explain the radio halo emission as originating
from the secondary pairs. Our cosmic ray limits are also
lower than the limits derived from \cite{Pfrommer2003} ($X_p<45\%\ldots25\%$ for
$\alpha_p=2.1\ldots2.5$)
which solely relied on the EGRET upper limit constraint.
For the case
$\alpha_p=2.1$ and $B=0.68\mu$G we find approximative equipartition between
particles and fields with $X_p\approx 0.05\%$.
Except for proton spectra harder than
$\alpha_p\leq 2.3$ we find in all cases
the radiation spectra at $>1$~keV from the secondary pairs to
lie below the corresponding photon spectra from the primaries. This is shown in Fig.~2
for $B=0.1\mu$G, which simultanously gives the most optimistic flux predictions
at high energies.
Below the hard X-ray band IC from both, primaries and secondary pairs determine the shape
of the volume-averaged spectrum. Depending on the proton spectral index
and overall hadron content in Coma, a turnover from primaries' to secondaries' dominated IC
below the soft X-ray band may occur. This is in agreement with the finding of \cite{Bowyer1998} that
the non-thermal halo component detected with the EUVE may stem from an additional component
of low-energy cosmic ray e$^-$ which we interpret as the secondary pairs.
Independent hints for a EUV emission of secondary pair origin has just been given
by \cite{Bowyer2004} who found a striking spatial correlation between the EUVE excess and
ROSAT thermal hard X-ray flux based on a re-examination of the EUVE data.
So far the EUVE excess radiation, if considered to be of non-thermal origin, has either been interpreted
as IC emission from low energy relic e$^-$ (\cite{Sarazin1999a}, \cite{Atoyan2000}) or explained by a
spectral break between the EUVE and HXR radiating e$^-$ (induced by a certain particle injection scenario), while a
secondary pair origin had erroneously been ruled out
(see \cite{Bowyer2004} for a discussion).

In the GLAST energy range non-thermal bremsstrahlung, followed by $\pi^0$-decay $\gamma$-rays
above $\sim 0.1$~GeV, will dominate, similar to the predictions given in \cite{Sarazin1999b}. Only for
hard input proton spectra IC radiation from the secondary pairs
will determine the GeV radiation. No $\gamma$-ray emission above $\sim$10 GeV is expected
for proton injection spectra as steep as $\alpha_p=5$.

\section{Detectability with gamma ray instruments}
Advances in the spatial
and spectral resolving capabilities of current/future high energy instruments
suggest to study the Coma cluster
continuum emission at energies higher than
the hard X-rays.
A detection at $\gamma$-rays may clarify on the spectral extent of the cluster's non-thermal emission,
may provide constraints on the acceleration processes realized in Coma, and also yields more sensible estimates
of a galaxy cluster contribution to the extragalactic $\gamma$-ray background (\cite{Berrington2003}, \cite{Fujita2003},
\cite{Gabici2003}, \cite{Miniati2002}). Observationally upper limits are currently provided by OSSE (\cite{Rephaeli1994})
and EGRET (\cite{Reimer2003}). INTEGRAL with its moderate continuum sensitivity\footnote{
http://www.rssd.esa.int/Integral/AO2/} permits a chance to detect
Coma as a marginally extended source up to a few 100 keV by ISGRI (\cite{Goldoni2001}). PICsIT and SPI, however, will not be able to detect
the Coma cluster given realistic observation times of $<10^7$sec. The current generation of imaging
Cherenkov telescopes (IACTS; \cite{Weekes2002}), in particular at northern locations, will reach
the required sensitivity only if significantly more than 50~hrs of observation will be accumulated.
Even then, the excellent resolving capabilities of IACTs
cannot be used to its full advantage due to the extended character of Coma's emission
where IACTs have a reduced sensitivity. This applies in particular to cases where
the dominant sub-GeV/TeV-emission component is originating from the outskirts, e.g.,
due to accretion shocks (\cite{Gabici2004}).

AGILE\footnote{http://agile.mi.iasf.cnr.it/Homepage/performances.shtml}, expected to have a similar performance as EGRET, might be able to verify the EGRET upper
limit. It is
the Large Area Telescope (LAT)
\footnote{http://www-glast.slac.stanford.edu/software/IS/glast\_lat\_performance.htm}, the main instrument
aboard GLAST, that has a realistic chance
to finally
detect Coma in continuum $\gamma$-rays. With its significantly better spectral and spatial resolution,
and up two orders of
magnitude improved sensitivity compared to EGRET, the $\pi^0$-decay component will be within
reach of LAT. Due to the similar spatial extent of Coma C and LAT's point spread function at GeV energies,
spatially resolved spectral information is difficult to gain. Although being photon-limited, LAT will benefit
from its wide field-of-view,
that allows a steady accumulation of exposure throughout
the expected mission life time for any observable object in the sky, inclusive
the Coma cluster.

\section{Conclusions}

The present work considers the role of the recently confirmed steepening of Coma's radio halo spectrum
in the GHz band for predicted fluxes in the high energy regime. Indeed, we found that the steepening
radio spectrum efficiently constrains the amount of hadronic cosmic rays through the
radiation channel of the secondary pairs produced in the decay chain of the hadronically produced
charged mesons. The implied upper limits for the hadronic cosmic ray energy density
range from 0.01\%\ldots 28\% of the thermal energy density, depending on the magnetic field
($B=0.1\mu\mbox{G}\ldots 2\mu$G) and proton injection spectral index ($\alpha_p=2.1\ldots 2.5$),
and are smaller than those used by other works. This might have severe implications
for the evolution of galaxy clusters, acting acceleration scenarios in cluster of galaxies
and the origin (secondary versus primary electron scenario) of Coma's radio halo.

Below the soft X-ray band we found that a turnover from primaries' to secondaries' dominated IC
emission may occur, depending on $\alpha_p$ and the hadronic cosmic ray content in Coma.
This is in agreement with the suggestions of \cite{Bowyer1998}
that the non-thermal halo component detected with the EUVE may stem from an additional population
of low-energy cosmic ray electrons which could in this scenario be interpreted as the secondary
pair component.
Independent hints for a EUV emission of secondary pair origin has just been given
by \cite{Bowyer2004} on the basis of a spatial correlation analysis between the EUVE excess and
the ROSAT thermal hard X-ray flux.

The steepening of the GHz radio spectrum leads to a decline of the IC and bremsstrahlung component
of the $\gamma$-ray spectrum already at 1-10 MeV and a few GeV, respectively, depending on the
magnetic field. We have shown that
the current continuum sensitivity of INTEGRAL's ISGRI at $>$ a few 100~keV for a $10^6$~sec observation
is insufficient to detect even the most optimistic predicted flux from Coma. The situation is even worse for
PICsIT and SPI.

$\pi^0$-decay $\gamma$-rays may extend Coma's $\gamma$-ray spectrum to TeV energies.
However, significant limits to its absolute flux are imposed by the radio spectrum (see above).
This leads to flux limits that are below
the point source minimum flux after 50~hrs on-source observations reached by modern generation
northern hemisphere Cherenkov telescopes like MAGIC and VERITAS. The case is even more hopeless for extended sources.

All predictions presented here are based on the assumption of power-law proton spectra in the Coma cluster.
Curved proton spectra
may be possible as a result of re-acceleration of the confined cosmic ray hadrons
in clusters of galaxies (\cite{GabiciNonPL}),
and this might lead to corresponding changes in the predicted limits.

It will be LAT of the GLAST mission that might finally be able to detect Coma in the $\gamma$-ray band if the magnetic
field and/or Coma's hadronic energy content is favorable.

\begin{acknowledgements}
AR's research is funded by DESY-HS, project 05CH1PCA/6, OR's by DLR QV0002. We thank the referee,
P. Blasi, for his constructive comments.
\end{acknowledgements}

\end{document}